\documentclass[sigconf,anonymous=false]{acmart}
\usepackage{subcaption}
\usepackage{graphicx}

\usepackage{enumitem}
\setlist{leftmargin=10pt}

\usepackage{capt-of}
\usepackage{booktabs}
\usepackage{varwidth}

\AtBeginDocument{%
  \providecommand\BibTeX{{%
    \normalfont B\kern-0.5em{\scshape i\kern-0.25em b}\kern-0.8em\TeX}}}

\copyrightyear{2022} 
\acmYear{2022} 
\setcopyright{acmlicensed}\acmConference[XXXX '22]{Proceedings of the 26th XXXX}{December 15--16, 2022}{Minneapolis, MN, USA}
\acmBooktitle{Proceedings of the  26th XXXX, December 15--16, 2022, Minneapolis, MN, USA}
\acmPrice{15.00}
\acmDOI{xxxxxx}
\acmISBN{999-9-9999-9999-9/99/99}

\begin{document}

\title[Transformer-based Pseudo-Relevance Feedback Model]{TPRF: A Transformer-based Pseudo-Relevance Feedback Model for Efficient and Effective Retrieval}


\author{Hang Li*}
\email{hang.li@uq.edu.au}
\orcid{0000-0002-5317-7227}
\affiliation{%
  \institution{University of Queensland}
  \city{Brisbane}
  \country{Australia}
}

\author{Chuting Yu*}
\email{chutingyu.cs@gmail.com}
\affiliation{%
  \institution{University of Queensland}
  \city{Brisbane}
  \country{Australia}
}

\author{Ahmed Mourad}
\email{a.mourad@uq.edu.au}
\orcid{0000-0002-9423-9404}
\affiliation{%
  \institution{University of Queensland}
  \city{Brisbane}
  \country{Australia}
}

\author{Bevan Koopman}
\email{bevan.koopman@csiro.au}
\orcid{0000-0001-5577-3391}
\affiliation{%
  \institution{CSIRO}
  \city{Brisbane}
  \country{Australia}
}

\author{Guido Zuccon}
\email{g.zuccon@uq.edu.au}
\orcid{0000-0003-0271-5563}
\affiliation{%
  \institution{University of Queensland}
  \city{Brisbane}
  \country{Australia}
}
\thanks{*Both authors contributed equally to this work.}
\renewcommand{\shortauthors}{Yu et al.}

\begin{abstract}
	
This paper considers Pseudo-Relevance Feedback (PRF) methods for dense retrievers in a resource constrained environment such as that of cheap cloud instances or embedded systems (e.g., smartphones and smartwatches), where memory and CPU are limited and GPUs are not present. For this, we propose a transformer-based PRF method (TPRF), which has a much smaller memory footprint and faster inference time compared to other deep language models that employ PRF mechanisms, with a marginal effectiveness loss. TPRF learns how to effectively combine the relevance feedback signals from dense passage representations. Specifically, TPRF provides a mechanism for modelling relationships and weights between the query and the relevance feedback signals. The method is agnostic to the specific dense representation used and thus can be generally applied to any dense retriever. 

\end{abstract}

\maketitle

\section{Introduction}

Pseudo-relevance feedback (PRF) aims to reduce the effect of query-passage vocabulary mismatch and thus improve search effectiveness by modifying the original query using information from top-ranked passages. PRF has been widely investigated in the context of traditional bag-of-words retrieval models~\cite{azad2019query}. 

Transformer~\cite{vaswani2017attention} based pre-trained deep language models~\cite{devlin2018bert,yang2019xlnet,dai2019transformer,radford2019language,raffel2019exploring} have shown promising results and improvements in information retrieval related tasks, and in particular for the cross-encoder architecture~\cite{lin2020pretrained,guo2020deep}. PRF methods for this architecture have also been proposed~\cite{padaki2020rethinking,zheng-etal-2020-bert,zheng2021contextualized,wang2020pseudo,yu2021pgt}. 
However, cross-encoder methods are computationally expensive, requiring both  query and passage encoding at query time. This limitation also applies to PRF methods for this architecture.

To overcome the computational cost limitation of cross-encoders, an alternative transformer based architecture has been devised: the bi-encoders, used by dense retrievers~\cite{zhan2020repbert,xiong2020approximate,lin2020distilling,lin2021batch,hofstatter2020improving,hofstatter2021efficiently,reimers2019sentence}. In such an architecture, the passage encoder and the query encoder are trained separately, passages are encoded at indexing time, and thus at inference time (i.e. querying time) only the query requires encoding. Empirically, dense retrievers achieve similar or even better effectiveness compared tp cross-encoder methods but with much higher efficiency~\cite{ding2020rocketqa,ren2021rocketqav2}.

The use of the PRF process in the context of dense retrievers has seen some initial  work~\cite{wang2021pseudo,yu2021improving,li2021improving}, but it is still a problem that is largely unexplored. Generally speaking, these methods consider the signal from the PRF passages in combination with the original query, input into a BERT-based query encoder fine-tuned for the PRF setting, and produce a new query encoding which is then matched against passages via dot product to generate a final ranking. For example, a current state-of-the-art PRF techniques for dense retrievers~\cite{yu2021improving,li2021improving}, known as ANCE-PRF~\cite{yu2021improving} in its setting that builds on top of the ANCE dense retriever~\cite{xiong2020approximate}, concatenates the text of query and feedback passages and provides them as input to an ANCE-based query encoder that has been fine-tuned for the PRF setting. Other dense retriever PRF methods differ in, among other aspects, the dense encoder used (e.g., ColBERT-PRF~\cite{wang2021pseudo}) and the use of dense vectors as input to the dense retriever encoder in place of text (Vector-based PRF~\cite{li2021pseudo}). Common drawbacks of these PRF methods are that: (1) the encoder used for the PRF signal is based on large pre-trained language models like BERT (for ColBERT-PRF, RepBERT-PRF~\cite{li2021improving}) and RoBERTa (for ANCE-PRF), rendering this encoder quite large in size (e.g., 503.4 MB for ANCE-PRF); 
and (2) the PRF signal given as input to the PRF encoder is often represented as text and thus requires expensive BERT inference operations to perform the PRF process, especially if CPUs, rather than GPUs, are used. Recall that the computational cost of a BERT inference grows as the size of the input grows. Thus the encoding of the query text plus the text of several passages is far more time consuming than the encoding the query alone.

In this paper, we consider the problem of devising PRF techniques for dense retrievers for systems with limited memory and computational resources, like cheap cloud instances or smartphones and smartwatches, while improving the effectiveness of the dense retriever to which PRF is applied and without adding considerable query latency. These small cloud instances and embedded systems are typically not equipped with GPUs and their memory is limited (e.g., 8GB for an Apple Watch and 1GB for a \$0.0104/hour AWS T3.micro instance). In these settings, the use of current PRF methods for dense retrievers is often hampered by the model size or the high query latency on CPUs, and one may be prepared to forgo some of the effectiveness that these state-of-the-art PRF methods can deliver for smaller and quicker, and thus cheaper, models (but that still improve on the baseline dense retriever). 

To tackle the challenges presented in this resource-constrained environment, we propose a transformer-based PRF method, called TPRF, which learns how to effectively combine the relevance feedback signal from dense passage representations without relying on PRF encoders based on large pre-trained language models.
Specifically, TPRF provides a mechanism for modelling the relationships and the weights between the query and the relevance feedback signals. To address the efficiency issue, in place of a large pre-trained transformer-based language model PRF encoder, we use a much smaller transformer with fewer layers and attention heads. With our specifically designed training regime, the loss in effectiveness compared to larger and more costly PRF methods for dense retrievers is often marginal. 
This research directly addresses the drawbacks discussed above. 



\section{Methods}
Next we describe our Transformer-based PRF (TPRF) method. TPRF addresses the limitations of  current PRF models for dense retrievers in terms of model size and efficiency by replacing the use of pre-trained deep language models as feedback signal encoders with purposely built transformer models. This is achieved by considering dense representations, rather than text, as the relevance feedback input signal.


\vspace{-6pt}
\subsection{Transformer-based PRF}

The TPRF method consists of a dense retriever, a transformer, and a purposely devised training strategy for the transformer. 

The dense retriever is responsible for producing a dense vector representation for each passage in the corpus. These are produced at indexing time and stored in a data structure that supports fast Approximate Nearest Neighbor (ANN) search, like a Faiss index. The dense retriever also produces a dense encoding of the initial query and a first ranking of the passages in answer to the query. The TPRF method does not prescribe which dense retriever need to be used -- and it is in fact agnostic to the dense retriever used. In the experiments in this paper we demonstrate TPRF in conjunction with the dense retrievers ANCE~\cite{xiong2020approximate} (and we indicate this method as ANCE-TPRF) and directly compare it against the ANCE-PRF method~\cite{yu2021improving}.

Once the first round of retrieval has been performed, the dense representations (and not the text) of the top-$k$ passages are used as feedback signal (i.e., PRF depth = $k$). The dense representation of the query is then stacked with the representations of the feedback signal to form a ($k+1, 768$) matrix. This matrix is provided as input to the transformer at the core of the TPRF model. The output is a $1,768$ dimensions dense vector representation. This output is used as the new query representation  from the PRF mechanism and used to perform a second round of retrieval using the original dense vector index: its result is the final ranking produced by TPRF.


At the core of TPRF is the transformer that is used to convert the input query representation and feedback passages representations into a new representation of the query with feedback. 
We choose the vanilla transformer encoder layer  proposed by~\citet{vaswani2017attention} to build the PRF query encoder in TPRF. Similar to other transformer layers used in language modelling tasks, the transformer encoder used in TPRF also leverages the self-attention mechanism to transform a sequence of input embeddings into contextualized embeddings. In our context, the transformer converts the query representation and the feedback passages representations into a new query representation. We inherit the key components of the vanilla transformer encoder layer, namely the positional encoding and the multi-head attention layer, to form the transformer encoder layers used in the TPRF query encoder. We apply the same positional encoding of~\citet{vaswani2017attention} on the input feedback passages embeddings to encode the rank position of the feedback passages returned from the first round of retrieval. The multi-head attention layer then produces the input query embedding and the feedback passage embeddings interact with each other and aggregate information via self-attention. As per common practice, the output of the multi-head attention layer follows a feed forward layer with residual connection and layer normalization.
Finally, we stack multiple transformer encoder layers to build the new query encoder. In our experiments, we study how different numbers of heads in the multi-head attention layers and different numbers of transformer encoder layers affect the effectiveness of TPRF.

\subsection{Training TPRF with Hard Negative Sampling}

At training time, after the initial retrieval using the baseline dense retriever, we pick the top-$k$ passages as the PRF, and we randomly sample 20 negatives among the passages that rank between position 10 and 200 in the initial ranking produced by the dense retriever. This sampling strategy was also used by~\citet{yu2021improving} for ANCE-PRF. We sample one target passage vector from the labelled relevant passages from the training dataset (MS MARCO training set queries, see Section~\ref{sec:settings}). 

For each training example, we randomly sample one positive passage $p^+$ and a set of negative passages $\{p_{1}^-, p_{2}^-, ..., p_{m-1}^-\}$ from the top passages retrieved by dense retriever; these labelled passages train the TPRF transformer with a supervised cross-entropy loss (Eq.~\ref{eq:loss}); the similarities between query, positive (target) passage, and negative passages are calculated using dot-product:
\vspace{-2pt}
\begin{equation}
\mathcal{L}_{CE} = -\log \frac{e^{s(q,p^+)}}{e^{s(q,p^+)} + \sum_{p^-}e^{s(q,p^-)}}
\label{eq:loss}
\end{equation}
\vspace{-2pt}
where $s(q,p)$ is the dot product of a query embedding $q$ and a passage embedding $p$.

\subsection{Theoretical Advantages of TPRF} 
We highlight that our TPRF uses a vanilla transformer model, i.e. the model is not pre-trained on a different task and data before being trained for PRF. This is unlike methods such as ANCE-PRF and its variants~\cite{yu2021improving,li2021improving}, that rely on pre-trained language models (RoBERTa) and the already fine-tuned ANCE encoder. This means the transformer in the TPRF model can be kept quite compact in size (number of parameters), thus resulting in a model that can potentially require less memory and less encoding time. Another advantage of TPRF is that its input are dense vectors, not text: this means that TPRF is not limited in the size of the input signal, i.e. the number of feedback passages. 
This is unlike ANCE-PRF and its variants, for which the feedback signal consists of the text of the top passages and thus is bounded by the maximum input size of the pre-trained language model.

\vspace{-4px}
\section{Experimental Setup} \label{sec:settings}


\subsection{Datasets and Evaluation Metrics}

We experiment with TREC Deep Learning 2019~\cite{craswell2020overview}, and 2020~\cite{craswell2021overview} (TREC DL 2019, TREC DL 2020). These datasets share the same passage corpus (MS MARCO \cite{nguyen2016ms}), which consists of $\approx$ 8.8M passages crawled by the Bing search engine, but with different queries (43 queries for 2019 and 54 for 2020) and graded relevance assessments.
We do not use the MS MARCO queries for evaluation. These queries are characterised by shallow judgements: on average only one labelled (and relevant) passage per query; no non-relevant labels are provided and thus is unsuitable to evaluate recall, which is often the target of improvement for PRF methods.
We instead use queries in the MS MARCO train set for training the considered methods. 

%
%
%
%


As for evaluation metrics, we use Average Precision (MAP), Reciprocal Rank (RR), Recall@1000 (R@1000), and normalised Discounted Cumulative Gain at various rank cut-off (nDCG@{1,3,10,100}). 
For all evaluation metrics, statistical significance testing using a paired two-tailed t-test with Bonferroni correction is performed. 

Along with effectiveness, we also measure the efficiency in terms of query latency of all methods in a CPU environment consisting of a consumer-grade Apple computer witg 2.4GHz 8-Core Intel Core i9 CPU and 64GB 2667MHz DDR4 memory. For measuring query latency, we randomly sampled 100 queries from the MS MARCO dev set, then issue them to each method, and report the average query latency measured. Finally, we also record the dimensions of each models in MB.


\vspace{-6px}
\subsection{Models}
For comparison, we report the effectiveness of the ANCE dense retriever~\cite{xiong2020approximate}; we also use ANCE to implement the ANCE-PRF method and we use ANCE also as the first stage model and source of dense representations for our ANCE-TPRF. For ANCE and ANCE-PRF we use the checkpoints provided by the corresponding authors; we also use Faiss~\cite{johnson2019billion} to build the dense index.

\vspace{-4px}
\subsection{TPRF Implementation and Training}
We implemented the vanilla transformer model in TPRF using Pytorch, Pytorch Lightning, and the Huggingface transformers library~\cite{wolf2020transformers} and separately combine TPRF with ANCE.
During the training process, we freeze the original query and passage encoders of ANCE: the query and passage representations generated from ANCE are not changed. 
We investigate different values of transformer layers (6, 8, 10, 12) and attention heads (4, 6, 12). 
We additionally also train a transformer with just one layer and one head, as the smallest possible instance of TPRF.
The hidden layer is fixed to 1024, input dimension to 768, the dropout to 0.2, and the learning rate to 1e-5. We set the batch size to 512, and trained for 50 epochs with AdamW optimizer, selecting the model with the highest nDCG@10 (following ANCE-PRF) on MS MARCO dev set, which was used for validation. 
All models were trained on a single NVIDIA Tesla SMX2 32GB GPU; training each model took approximately 26 hours for 50 epochs. 
For comparison, we execute all PRF methods (ANCE-PRF and ANCE-TPRF) with PRF depth $k=3$: this is the best configuration for ANCE-PRF and the checkpoint distributed for this model refers to $k=3$. We highlight that the effectiveness of ANCE-TPRF may also vary largely depending on $k$; we leave this investigation to future work.

\begin{table*}[ht!]
	\centering
	\resizebox{!}{0.18 \columnwidth}{%

\begin{tabular}{lrrrr|rrrr}
	\toprule
	&                             \multicolumn{4}{c}{TREC DL 2019}                              &                                    \multicolumn{4}{c}{TREC DL 2020}                                     \\
	&               MAP &            R@1000 &        MRR &    nDCG@10 &               MAP &            R@1000 &        MRR &            nDCG@10 \\ \midrule
	$^a$ ANCE      &            0.3710 &            0.7554 &     0.8372 &       0.6452 &            0.4076 &            0.7764 &     0.7907  &            0.6458 \\
	$^p$ ANCE-PRF  & \bf 0.4253$^{+a}$ & \bf 0.7912$^{+a}$ & \bf 0.8492 &  \bf 0.6807 & \bf 0.4452$^{+a}$ & \bf 0.8148$^{+a}$ & \bf 0.8371$^{+a}$ & \bf 0.6948$^{+a}$ \\
	$^t$ TPRF-ANCE &     0.3855$^{-p}$ &     0.7771$^{+a}$ &     0.8358 &      0.6578 &     0.4135$^{-p}$ &            0.7865 &     0.7743$^{-p}$ &       0.6407$^{-p}$ \\ \bottomrule& & 
\end{tabular}

	}
	\caption{Effectiveness of dense PRF methods; bold values are the best for that metric and dataset. Statistical significant differences ($p<0.05$) are indicated using the superscript character corresponding to each method.}
	\label{tbl:effectiveness}
\end{table*}

\section{Results}\label{sec:results}



\begin{figure*}[t]
	\centering
	\includegraphics[width=\textwidth]{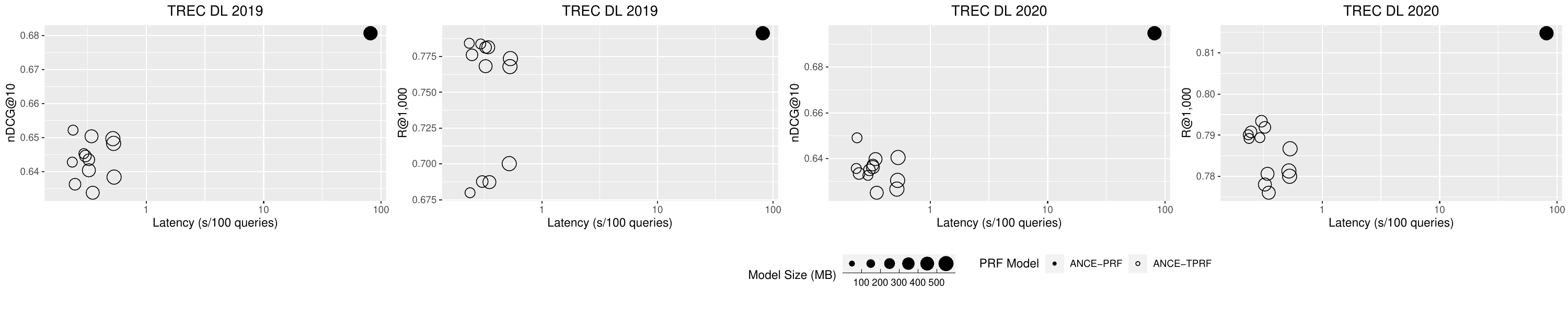}
	\caption{Relationship between effectiveness (measured as nDCG@10 and R@1000), query latency and model size for PRF methods, across datasets. For ANCE-TPRF, we display different configurations (w.r.t. number of layers and attention heads). } \label{fig:trade-offs}
\end{figure*}

\subsection{Overall Effectiveness}
Table~\ref{tbl:effectiveness} reports the effectiveness of ANCE, ANCE-PRF and ANCE-TPRF on TREC DL 2019 and 2020. For ANCE-TPRF we report the model configuration that obtained the highest nDCG@10 on the validation set: this is the model with 6 transformer layers and 12 attention heads.

From the results, we observe that ANCE-TPRF always outperforms ANCE except in nDCG@3 for DL 2019 (only $0.3\%$ loss) and in MRR and nDCG@k in DL 2020. These differences are however not statistically significant.  Differences between ANCE-PRF and ANCE-TPRF are more substantial. In particular, ANCE-TPRF is consistently outperformed by ANCE-PRF and some of the differences are significant. If one was interested in effectiveness alone, then it is clear ANCE-PRF is a better method, and that ANCE-TPRF does provide some gains over ANCE especially for metrics that value some form of recall (R@1000, MAP).

\subsection{Trade-off between Effectiveness, Model Size and Query Latency}
The motivation for proposing TPRF, however, was to achieve a balance between the improvements in effectiveness {pro\-mi\-sed} by PRF and the need for efficiency (in query latency and memory consumption) required by the resource constraint environment we are interested in.
Next, then, we study the trade-offs between effectiveness, query latency and model size. This analysis is reported in Figure~\ref{fig:trade-offs}. For TPRF-ANCE, we report the best checkpoint (according to nDCG@10 on validation) for each configuration of number of transformer layers and number of attention heads. From the figure we can make the following observations:

\begin{itemize}
	\item Concerning the variation in effectiveness across ANCE-TPRF models, we note that their differences are not large for nDCG@10 across both datasets: there is no statistical significant difference between the best and the worst performing model. The differences in terms of Recall@1000 are more marked, especially for DL 2019, where (1) models seem cluster into two groups in terms of effectiveness, (2) differences between the best and the worst model are significant.
	
	\item ANCE-PRF consistently outperforms ANCE-TPRF across metrics and datasets, with the exception of Recall@1000 on DL 2019, for which differences are minor. Despite this, differences between the best ANCE-TPRF model and ANCE-PRF are not significant for DL 2019; and they are only significant in DL 2020 for nDCG@10.
	
	\item However, the larger effectiveness of ANCE-PRF comes at a significant cost: higher query latency. In fact, ANCE-TPRF is blazing fast on a commodity CPU, with 100 queries taking a fraction of a second to run. ANCE-PRF instead requires in the proximity of 100 seconds to run the same amount of queries.
	
	 \item In terms of model size, it is interesting that for ANCE-TPRF, model size and effectiveness are not correlated. In fact, it is often the case that the best models span different sizes, and that small models perform quite well and at times better than larger models -- for example, the ANCE-TPRF model with highest nDCG@10 on DL 2020 (0.6491) has a size of 299.2 MB, while the largest model has size 582.9 MB and achieves a nDCG@10 of 0.5581. 
	 
	 \item The ANCE-TPRF models that are closest in effectiveness to ANCE-PRF exhibit less memory consumption than ANCE-PRF. The size of ANCE-PRF is 503.4 MB. The best ANCE-TPRF model for DL 2019 on both nDCG@10 and R@1000 is 299.2 MB; the best ANCE-TPRF model for DL 2020 is 299.2 MB for nDCG@10 and 393.8 MB for R@1000.
	 
	 \item For TPRF, model size is determined by the number of transformer layers $l$, rather than that of attention heads $h$. For example, the smallest models are 299.2 MB in size and are obtained with the combinations $(l=6, h=4)$, $(l=6, h=6)$ and $(l=6, h=12)$. 
	 
	 \item The TPRF models with less memory footprint are also the ones with the lowest query latency. Although the query latency of the smallest models (0.235 s/100 queries, 299.2 MB) is almost half that of the the largest (0.533 s/100 queries, 582.9 MB), the absolute difference is modest ($\approx 0.30$), compared to the query latency of ANCE-PRF. 
	 
 \end{itemize}

The model with 1 transformer layer and 1 attention head, not included in the analysis above, deserves particular consideration. This ANCE-TPRF setting is characterised by the smallest transformer, at just 62.7 MB in size, and the smallest query latency, at 0.185 s/100 queries. Although not as complex as ANCE-PRF or the other ANCE-TPRF models, this model is still able to improve ANCE in terms of MAP and R@1000 on both datasets.

In summary, ANCE-PRF obtains the highest effectiveness across metrics and datasets, while ANCE-TPRF trades-off higher effectiveness for smaller and faster models.
This trade-off is especially useful in memory-constrained settings: one can train a smaller TPRF model, thus forgoing some effectiveness, but still can get improvements over the base dense retriever models.

\vspace{-3px}
\subsection{Query Latency and Scalability to PRF Depth}
In ANCE-PRF, query latency is determined by the length of the input (query text plus text of the appended top $k$ passages used as feedback): when the input length increases, the inference time also drastically increases. This is because inference speed is impacted bu the self-attention computation: this is an expensive process~\cite{dai2019transformer,vaswani2017attention} (in terms of runtime) and larger the length of the input, more self-attention computations are required.
The input of ANCE-PRF however is also bounded by the maximum size of the input to RoBERTA ($\approx 512$ tokens): ANCE-PRF's latency will never surpass a certain limit: in our experiments, this occurs for $k=5$ for which latency is $\approx125$ s/100 queries. This happens because any part of the input that surpasses the input size limit is truncated and thus ignored.

\begin{figure}[t]
	\centering
	\includegraphics[width=1\columnwidth]{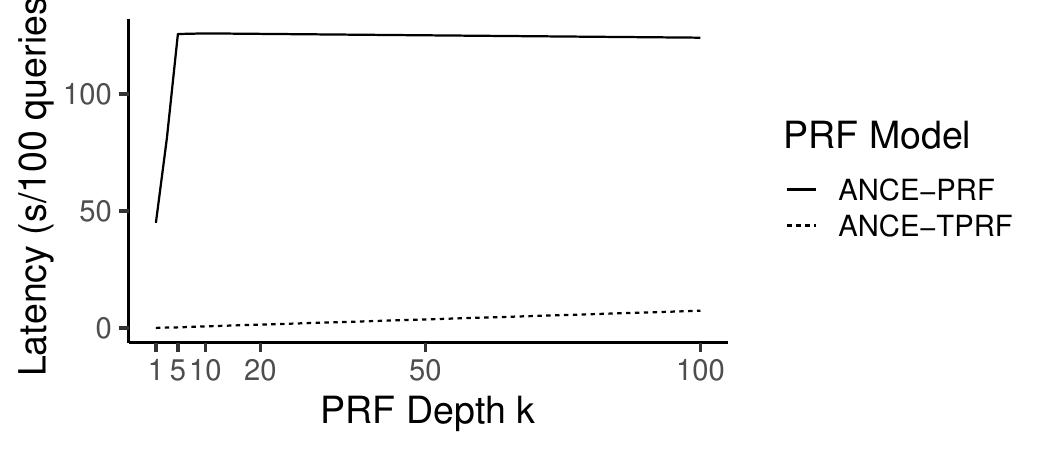}
	\captionof{figure}{Query latency for the PRF models as a function of the PRF depth $k$.}
	\label{fig:latency-PRFdepth}
\end{figure}

In ANCE-TPRF, query latency increases with the increase of $k$, i.e. the number of feedback passages. However, the TPRF inference time is substantially lower than that of ANCE-PRF, regardless of $k$. This is because, while in ANCE-PRF larger $k$ values mean more text to encode (so more time required), in ANCE-TPRF larger $k$ values simply mean more dense vectors been accumulated in the input -- and increases in this number do not determine increases of the transformer's inference time, which stays the same. Instead, as $k$ increases, the time taken to form the input matrix increases, but this is negligible. We demonstrate these remarks in Figure~\ref{fig:latency-PRFdepth}, where the PRF depth $k$ is varied up to 100. ANCE-PRF query latency increases very rapidly as $k$ increases, but then plateaus when $k=5$ because the input size saturates (i.e. exceeds the maximum of 512 tokens). The query latency of ANCE-TPRF also increases as $k$ increases, but this increase is very slow-paced; in addition ANCE-TPRF is not limiteed in input size. Even after $k=100$ passages have been given as input to ANCE-TPRF, its query latency is still largely less than the query latency of ANCE-PRF when $k=5$.

\section{Conclusion}

In this paper, we consider pseudo relevance feedback methods for dense retrievers in a resource constrained environment. Specifically, we observe that current PRF methods based on large pre-trained language models and fine-tuned dense retrievers query encoders are not suitable to resource constrained environments such as embedded systems with limited memory and no GPU. These models in fact are often very large, and their query latency on CPU can be considerable -- especially if the feedback signal is large or the CPU is not powerful.

To overcome these challenge we devise a transformer-based PRF approach, called TPRF, which leverages transformer encoder layers that do not use the feedback passage text as input, but their dense representations. This allows us to produce more compact models (i.e. smaller in size), and because they have less parameters they also guarantee a lower query latency, thus being usable in settings with low memory and low CPU capabilities. In addition, our TPRF method addresses some of the limitations of previous PRF models, such as the limited amount of PRF signal they can handle -- what is more, TPRF is highly scalable to increasing input signal. 
Empirical results in the context of the ANCE dense retriever confirm that our TPRF method is characterised by lower query latency and memory utilisation than other PRF methods, while having to forgo often only minimal effectiveness (i.e. no statistical significant losses).



%


\bibliographystyle{ACM-Reference-Format}
\bibliography{references}










\end{document}